\documentclass[mnras, twocolappendix]{emulateapj}
\usepackage[fleqn]{amsmath} 
\usepackage{amsfonts} 
\usepackage{amstext}
\usepackage{amssymb} 
\usepackage[colorlinks=true, citecolor=blue, linkcolor=blue,
breaklinks=true]{hyperref}
\usepackage{natbib}
\usepackage{graphicx}

\def\etal{\it et al. \rm }

\begin{document} 

\title{Colors of Dwarf Ellipticals from {\it GALEX} to {\it WISE}}

\author{James M. Schombert$^{A,B}$}
\affil{$^A$Department of Physics, University of Oregon, Eugene, OR USA 97403}
\affil{$^B$jschombe@uoregon.edu}

\begin{abstract}

\noindent Multi-color photometry is presented for a sample of 60 dwarf ellipticals
(dE) selected by morphology.  The sample uses data from $GALEX$, SDSS and {\it WISE}
to investigate the colors in filters $NUV$, $ugri$ and $W1$ (3.4$\mu$m).  We confirm
the blueward shift in the color-magnitude relation for dwarf ellipticals, compared to
CMR for bright ellipticals, as seen in previous studies.  However, we find the
deviation in color across the UV to near-IR for dE's is a strong signal of a younger
age for dwarf ellipticals, one that indicates decreasing mean age with lower stellar
mass. Lower mass dE's are found to have mean ages of 4 Gyrs and mean [Fe/H] values of
$-$1.2.  Age and metallicity increase to the most massive dE's with mean ages similar
to normal ellipticals (12 Gyrs) and their lowest metallicities ([Fe/H] $= -$0.3).
Deduced initial star formation rates for dE's, combined with their current metallicities
and central stellar densities, suggests a connection between field LSB dwarfs and
cluster dE's, where the cluster environment halts star formation for dE's triggering
a separate evolutionary path.

\end{abstract}

\section{Introduction}

Ellipticals represent one of the most carefully studied type of galaxies with respect
to stellar populations.  They exhibit the simplest morphological as
well as internal kinematics and, thus, are well-modeled as a single stellar
population rather than the kinematically distinct components found in disk galaxies.
They have the highest masses (i.e., luminosities) of all galaxy types and, therefore,
are the clearest signposts at high-redshift.  Studies of galaxy evolution often focus
on ellipticals owing to early indications that their spectrophotometric changes are the
simplest to model and trace, reliability, through cosmic time (so-called passive
evolution).  In addition, ellipticals are the only morphological type that
inhabits every galaxy environment from the richest clusters to the field, and present
a coherent appearance from the dwarf ellipticals to the brightest cluster members.

The study of stellar populations in ellipticals has historically taken three
different routes.  The first is the use of optical and near-IR colors to interpret
the integrated light of the underlying stellar population.  The discovery of the
color-magnitude relation (CMR, Sandage \& Visvanathan 1978), the separation of
morphology types by color (Tojeiro \etal 2013) and different galaxy components by
color (e.g., core versus envelope, Head \etal 2014) were often our earliest
explorations into the stellar populations and the meaning of color with respect to
the star formation history of galaxies (Tinsley 1978).  Technological improvements in
the 1980's led to a second, and obvious, extension of multi-color work through a
higher resolution of the spectroenergy distribution (SED) of galaxies with the focus
on various spectral indices related to different types of stars found in a stellar
population.  This type of investigation reached a peak with the development of the
Lick/IDS line-strength system (Worthey \etal 1994; Trager \etal 2000) where a set of
specific spectral features were shown to correlate with the two primary
characteristics of a stellar population, its age and mean metallicity.  Guided by SED
models of the Lick/IDS line-strength system (see Graves \& Schiavon 2008), these
spectral indices became the observable of choice to study nearby and distant galaxy
stellar populations.  Lastly, with the launch of HST, space imaging provide the best
study of a stellar population by direct examination of their color-magnitude
diagrams. The number of pure ellipticals open to HST CMD imaging is limited to the
nearby Universe and, therefore, for a majority of stellar population studies the
Lick/IDS line-strength system is the method of choice.

The use of optical and near-IR colors for investigating stellar populations still has
a useful role to play even in the spectroscopic era.  For example, higher
signal-to-noise is acquired for faint, distant galaxies using colors.  Large areal
surveys, such as those obtained by wide-field cameras using well chosen filter sets,
provide numerically superior datasets to spectroscopic surveys.  In the 1990's, Rakos
\& Schombert pioneered the use of narrow band filters selected to cover
age/metallicity features around the 4000\AA\ break as an fast and efficient system to
study cluster galaxies with direct imaging (see Rakos \& Schombert 1995).  The
results from those studies confirmed a passive evolution for the stellar populations
in cluster ellipticals, but was in sharp disagreement with the results from
spectroscopic surveys that found much younger ages and lower metallicities for the
same objects (Trager \etal 2000; Graves \etal 2009; Conroy \etal 2014).  Larger SDSS
samples (Bernardi \etal 2005, Gallazzi \etal 2005; Graves, Faber \& Schiavon 2010) presented a more
diverse range of ages and metallicities, and more sophisticated analysis techniques
(see Johansson \etal 2012; Conroy \etal 2014; Worthey \etal 2014), all of which
reinforced the conclusion of younger ages and lower metallicities with decreasing
luminosity and stellar mass for bright ellipticals.

One deficiency in spectroscopic surveys is that they are less able to explore low
luminosity/mass realm of ellipticals, the dwarf ellipticals.  Their combined low
absolute luminosities and depressed central surface brightnesses made spectroscopic
observations of dwarf ellipticals a particular challenge even for nearby clusters
and, in practice, impossible for high-redshift systems.  Given the various scenarios
where bright ellipticals are constructed from low mass systems by hierarchical
mergers (Driver 2010), a deficiency in accurate stellar populations values (e.g., age
and metallicity) for dwarf ellipticals is a continuing problem in comparing
observations with theory.

This paper attempts to extend our knowledge of stellar populations in low mass
systems by presenting a comprehensive analysis of dwarf elliptical colors using
archival data from the near-UV ({\it GALEX NUV}) to the far-IR ({\it WISE} 3.4$\mu$m)
in an effort to extend our estimates of age and metallicity to the low mass realm.
This study is an extension of the color analysis of bright ellipticals ($M_g < -20$)
from Schombert (2016) that focuses on presenting a large sample of morphologically
classified dwarf ellipticals (class dE) and bridging the gap between normal and dwarf
ellipticals (the luminosity range between $-$18 and $-$20).  As demonstrated in
Schombert (2017), this gap is populated by the rare class of faint ellipticals with
power-law surface brightness profiles, in the same structural family as normal
ellipticals.  The family of dE's have structurally distinct profiles from normal
ellipticals, yet appear to transition coherently in color with normal ellipticals
(Caldwell 1983).

This study will provide a detailed comparison of dwarf ellipticals with previous
color studies and will also examine the differences between various photometric
relationships.  The colors presented herein will anchor the zeropoint of bright and
dwarf elliptical colors for use by high-redshift studies.  In addition, these colors will also
provide a window into the stellar populations of ellipticals by comparison with
simple and multi-metallicity population models where the wide wavelength coverage
offers an avenue to break the age-metallicity degeneracy that plagues optical colors
(Worthey 1994).  

\section{Sample}

The bright elliptical data for this study ($M_g < -20$) were based on the sample
defined by Schombert \& Smith (2012), a purely morphological sample of ellipticals
selected from either the Revised Shapley-Ames (RSA, a catalog selected
by luminosity) and Uppsala Galaxy Catalog (UGC, an angular limit catalog).  The
sample was restricted to large angular-sized ($D > 2$ arcmins) galaxies and required
have been imaged by the {\it 2MASS} project (we used $J$ magnitudes as our baseline).
In addition, the sample had to satisfy a criterion of isolation from foreground or
background objects (i.e., there are no nearby bright stars or companion galaxies that
would distort the surface brightness isophotes).

The resulting $JHK$ surface photometry was presented in Schombert \& Smith (2012),
and the final sample consisted of 436 bright ellipticals.  That sample was then
cross-correlated with the {\it Galaxy Evolution Explorer} ({\it GALEX}, Martin \etal
2005), SDSS and {\it Spitzer} image libraries for existing data from 226.7nm ({\it
GALEX} $NUV$) to 3.6$\mu$m (channel 1, {\it Spitzer}).  Using automated scripts to
browse the various mission websites resulted in 2,925 image files from the four
missions.  Overall there were 436 ellipticals in the {\it 2MASS} sample, of which 149
had {\it GALEX} data, 252 had matching SDSS images and 149 with archived {\it
Spitzer} 3.6$\mu$m images.  These numbers are primary determined by the varying sky
completeness of each mission.

In the original sample, only 5\% of the galaxies were fainter than $M_g = -20$, which
sharply degrades the ability to study the color-magnitude relation to low
stellar masses.  To extend the elliptical sequence, we have collected 60 more
ellipticals from the recent early-type catalog of Dabringhausen \& Fellhauer (2016,
hereafter DF catalog) specifically for low absolute magnitude.  Again, pure
elliptical morphology and isolation were the requirement, plus the target had to
be in the SDSS DR13 image library.  In addition to these faint ellipticals (classed
as E in the DF catalog), 62 dwarf ellipticals were also selected from the dwarf
elliptical sample of Lisker, Grebel \& Binggeli (2008) for study.  Of this 62, 49 are
classed dE(N), eight are classed dE(nN) and five as dE(bc) based on the Lisker
scheme.  A majority of these galaxies are in the Virgo cluster.  The Virgo sample was
combined with a sample of group dE from the DF catalog for a total dwarf sample of 62
galaxies.  The combined sample (bright, faint and dwarf) contains 374 ellipticals
with, at least, photometry from SDSS $ugri$ images.

Each object in the total sample was also inspected for evidence of emission lines
(excluding AGN features), dust or other signatures of recent star formation when
spectroscopic data was available in the SDSS archive.  The idea here was to find a
sample of ellipticals that was as similar in terms of morphology and star formation
history as possible.  While some contamination of inner colors due to low level AGN
activity was acceptable, their effects had to be restricted to inside the various
mission's PSF or the galaxy was rejected from the sample.  All the images were
ellipse subtracted to look for asymmetric features that might be a signature of
recent mergers or dust lanes.  While the usual selection of boxy-like and disk-like
residuals were observed, there were no obvious linear features.   In addition, color
subtracted frames were examined for an evidence of dust lanes, none were detected in
the UV and optical images.

\subsection{Data Reduction}

Data reduction of the flattened, calibrated images from each mission was performed
with the galaxy photometry package ARCHANGEL (Schombert 2011).  These routines, mostly
written in Python, have their origin back to disk galaxy photometry from the late
1980's and blend in with the GASP package from that era (Cawson 1987).  The package
has four core algorithms that 1) aggressively clean and mask images, 2) fit
elliptical isophotes, 3) repair masked regions then perform elliptical aperture
photometry and 4) determine aperture colors and asymptotic magnitudes from curves of
growth and determine accurate errors based on image characteristics, such as the
quality of the sky value.

The photometric analysis of galaxies branches into four areas; 1) isophotal analysis
(the shape of the isophotes), 2) surface brightness determination and fitting (2D
images reduced to 1D luminosity profiles), 3) aperture luminosities (typically using
masked and repaired images and elliptical apertures) and 4) asymptotic or total
magnitudes (using curves of growth guided by surface brightness data for the halos,
see Schombert 2011).  Ellipticals are the simplest galaxies to reduce from 2D images
to 1D luminosity profiles since, to first order, they have uniformly elliptical
shaped isophotes (Jedrzejewski 1987).  Where many ellipticals display disky or boxy
isophotal shapes (Kormendy \& Bender 1996), this deviation is at the few percent
level and has a negligible effect on the surface brightness profile, aperture
luminosities or colors values (Schombert 2013).

Surface brightness determination and fitting consumes a large fraction of the
processing time for ellipticals.  Accurate surface brightness profiles require
detailed masking to remove foreground stars, background fainter galaxies and image
artifacts.  While cleaning an elliptical galaxy's image is simplified by the lack of
HII regions, dust lanes or other irregular features, the final accuracy of the
profile will be highly dependent on the quality of the data image, in particular the
flatness of the image and knowledge the true sky value.  The smooth elliptical shape
to early-type galaxies results in very low dispersions in intensity around each
isophote that, in turn, makes the removal of stellar and small background galaxies an
iterative task.

Following the prescription outlined in Schombert \& Smith (2012), we processed all
the mission images in the same manner.  Despite the differing plate scales (i.e.,
arcsecs per pixel), orientations on the sky and flux calibrations, the {\it GALEX},
SDSS, {\it 2MASS} and {\it Spitzer} missions all provide well flattened final data
products, usually free of any obvious artifacts.  Very little image preparation was
required, other than confirming that the targets in the images were, in fact, the
correct galaxy (galaxy misidentification in the archive servers was not uncommon).
This was accomplished by comparison with the PSS-II J images at STScI/MAST and crude
luminosity estimates compared to the RC3 magnitudes (de Vaucouleurs \etal 1991).

Isophotal fitting on each image begins with a quick visual inspection of the field
for manual suppression of artifacts and marking the center of the galaxy.  Then, an
iterative ellipse fitting routine begins outside the core region, moving outward
fitting the best least-squares ellipse to each radius until the isophote intensity
drops below 1\% of sky.  The routine then returns to the core to finish the inner
pixels in a like manner.  During the ellipse fitting, pixels greater than (or less
than) 3$\sigma$ from the mean intensity are masked and removed using a 50\% growth
radius.  The resulting fits are output as mean intensity, dispersion around the
ellipse, major axis, eccentricity, position angle (and errors) plus the first four
intensity moments.  Conversion to surface brightness profiles uses the generalized
radius, the square root of the major times minor axis ($\sqrt{ab}$).  All spatial
parameters will be quoted using the generalized radii.

Resulting elliptical isophotes are calibrated (intensity and pixel size) using the
standard pipeline calibrations provided by the missions, then processed into surface
brightness profiles.  Various fitting functions have been applied to elliptical
surface brightness profiles over the years.   A full discussion of their various
strengths and weaknesses can be found in Schombert (2013).  In brief, the S\'{e}rsic
$r^{1/n}$ provide the best fits over the full range of surface brightness, but
suffers from coupling between its shape and characteristic radii parameters that
reduce its usefulness.  Templates are stronger match to the shape of elliptical
profiles (Schombert 2015), but are only parametrized by luminosity and do not provide
any structural metrics.  In the end, we found that empirically determined parameters,
such as half-light radius ($r_h$) and mean surface brightness ($<\mu>$), are the most
strongly correlated parameters with luminosity or stellar mass.

As this study is primarily concerned with colors, the determination of luminosity in
a consistent and accurate manner from the datasets is of the highest priority.
Aperture luminosities are calculated using the ellipses determined by the isophote
routines.  Care was taken to make sure that the same eccentricities and position
angles were used across the various mission images.  None of the sample galaxies
display any variation in eccentricity or position angle at the 2\% level from the
near-UV to the far-IR.  Thus, aperture values were, in effect, determined using
fixed radii in kpcs.

The total luminosity of a galaxy is a much more problematic value to determine.  The
procedure used here is outlined in Schombert (2011), where elliptical apertures are
determined by a partial pixel routine from the masked images where the masked regions
have been filled by the local mean isophote.  While it seems obvious that masked
regions would reduce the calculated flux inside an aperture, in fact, this effect is
rarely more than 5 to 10\% the total flux of an elliptical.  However, this effect is
also unlike Poisson noise in that it always works to reduce the measured luminosity.  The
cleaned images are then integrated as a function of radius to produce curves of
growth.  An added feature is that as the outer aperture are integrated, their fluxes
are corrected by the mean isophotal intensity as given by either the raw surface
brightness profile or the S\'{e}rsic $r^{1/n}$ fit to the profile.  This reduces the
noise in the outer apertures and often produces a smoother convergence to a stable
total flux.

The half-light luminosity, and radius, used in our bright elliptical sample, was
found to be unstable for the dwarf ellipticals.  Their smaller angular size and
exponential profiles introduce unacceptable error in the determination of the
half-light radius.  Instead, for this paper, we have used the Holmberg radius which
is the radius where the $g$ surface brightness profile reaches 25 mag arcsecs$^{-2}$.
The Holmberg luminosity is the aperture luminosity calculated inside this radius and,
typically, results in 92\% of total of the total luminosity (although this varies
slightly with filter choice, Schombert 2016).  While this choice has little impact on
measured colors (color gradients are the dominate source of color variation), for
relationships that compare total luminosity (a proxy for total stellar mass) with
color (e.g., the CMR) then the aperture luminosity is corrected for the missing 8\%
to bring the luminosities in alignment with the half-light values from Schombert
(2016).

In a majority of the missions, the error quoted at the archives for the total and
aperture magnitudes severely underestimates the actual error found in this study (see
Schombert 2016, \S2.2, 2.3, 2.4 and 2.5).  This is due to the fact that their error
calculations focus, primarily, on the Poisson noise that is proportional to device
sensitivity and exposure time.  However, for large extended sources ($D > 2$ arcmins)
or ones that are low in mean surface brightness (i.e., dwarfs), the primary source of
noise is uncertainty in the sky value and the variation of sky across the image.  Sky
for this study was determined by two separate algorithms.  The first is the manual
section of between 10 and 20 sky boxes (typically 20 by 20 pixels in size) in regions
surrounding the target (outside its halo), but separate from other galaxies or bright
star halos.  The dispersion of the mean from averages in each sky box provide the
best value for the uncertainty on the sky value (Schombert 2013).  Total errors
quoted in this paper are then calculated as 3$\sigma$ from the mean sky value applied
to the sum of the pixels in each aperture.  As a check to the correct sky value, the
elliptical isophotes are fixed in shape and tabulated beyond the galaxy radius to the
edge of the frames.  These outer ellipse intensity values should converge on the sky
value determined from the sky boxes.  In the few cases where the two values disagree,
the ellipse sky value was used, but the error estimates continued to use the
dispersion between the sky boxes.

Full surface brightness analysis is performed on all the filter images.   Thus, a
second determination of color is possible by examining color as measured by the
difference in the surface brightness profiles directly.  While any particular color
surface brightness isophote contains much more error than the comparable aperture
color at the same radius, at large radii this method can be more informative as the
number of pixels used is large compared to the sky error.  Of course, color gradients
are the primary use for multi-filter surface brightness profiles.  And, again, with
the regularity of shape for ellipticals, the run of color with radius is a direct
measure of the projected 2D distribution of stellar populations.  The best measure of
the core color of an elliptical is the interpolation of the color surface brightness
profiles to zero radius.  

Apparent uncorrected luminosities are designated by lowercase letters (e.g.,
$m_{NUV}$ for raw {\it GALEX} $NUV$ luminosities).  Absolute luminosities, radii and
colors are corrected for Galactic extinction following the prescription of Cardelli,
Clayton \& Mathis (1989) and $E(B-V)$ values from NED.  Magnitude units varied
between the missions; {\it GALEX}, SDSS and {\it WISE} use the AB/Vega system
($NUV,u,g,r,i,W1,3.6$) and {\it 2MASS} used the Johnson system ($J,H,K$).  We have noted in
our discussions when the various units are used.  Distances used are the 3K CMB
distances using the Benchmark model values for the standard cosmological constants
(in particular, $H_o = 75$ km/sec/Mpc).  A majority of the dE's were located in the
Virgo cluster and a distance modulus of 31.09 was assumed.  As none of the galaxies
in this sample have redshifts greater than 0.04, no k-corrections were applied to the
data.  The final data products are too extensive to be listed in this publication.
Instead, the author maintains all the data, reduction scripts and log files at his
website (http://abyss.uoregon.edu/$\sim$js).  We follow the philosophy of presenting
all the reduction techniques as user enabled scripts, rather than a detailed
description of the various steps leading from raw images to final luminosities and
colors.

\subsection{{\it True Internal Colors}}

One point about galaxy colors that is often unspoken, and assumed to be known to the
reader, is that a color quoted for a particular region in a galaxy does not represent
the color of the stellar population at that 3D coordinate, but rather a pencil beam
average of the stellar population colors between the observer and infinity.  Thus,
the measured color at a particular elliptical annulus traced by an elliptical fit of
radius, $r_o$, is, in fact, the luminosity weighted sum of colors of all the
isophotes at radii $r > r_o$.  This effect will be greater for small radii, where the
pencil beam crosses through all the isophotes in the galaxy profile, and less for
outer radii, where the pencil beam passes through a fewer number of isophotes with
colors similar to the starting radii (for typical color gradients).  Negative
color gradients (decreasing color with radius) will result in underestimating
the true color of a position with more significance for stronger gradients and
steeper profiles.

\begin{figure}[!ht]
\centering
\includegraphics[scale=0.38,angle=0]{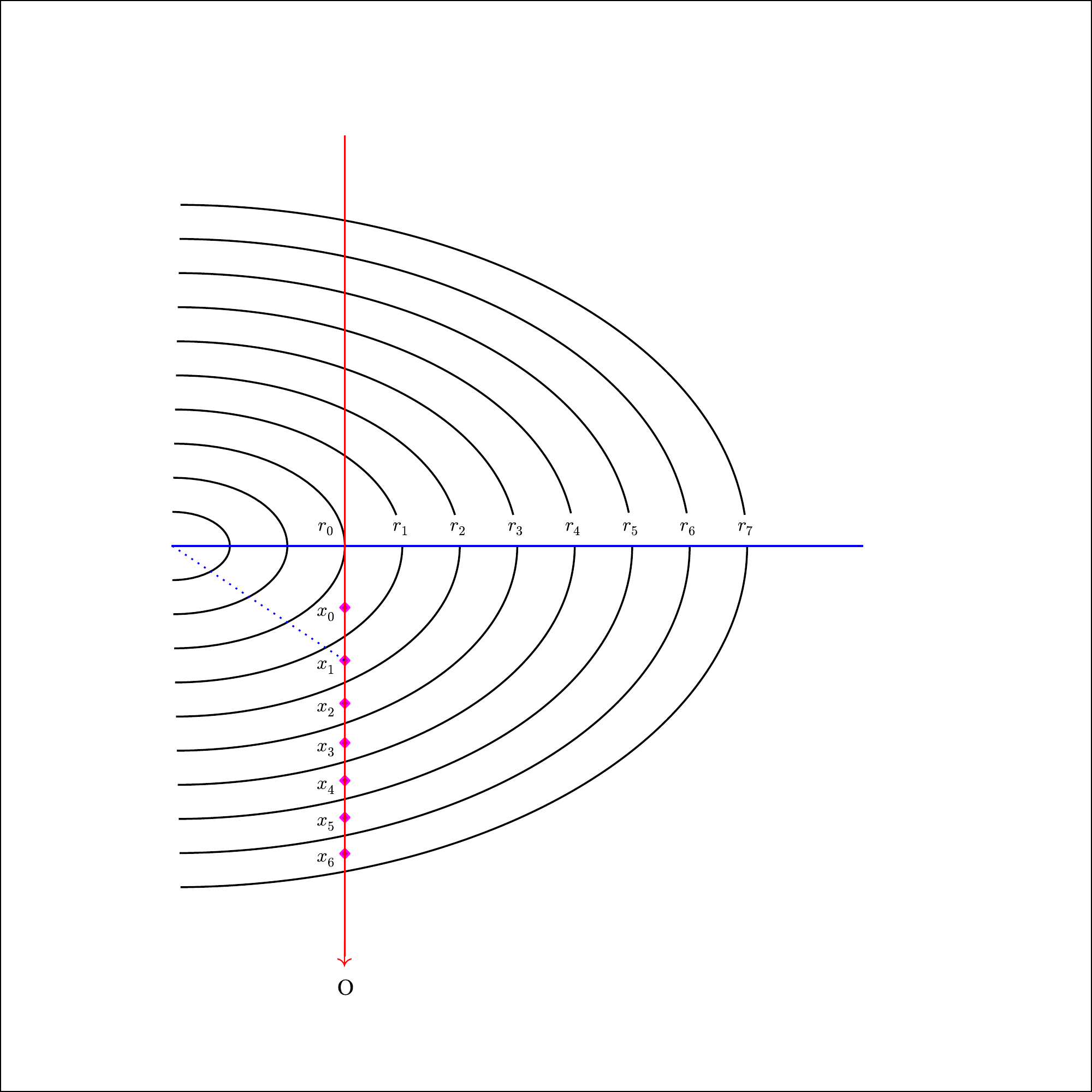}
\caption{\small Calculation of the true color of a 3D section of an elliptical.  Any
measurement of a color value at a projected radius ($r$) is, in fact, the sum of the
color intensities through a cross section (red line) toward the observer (O).  Two
factors determine the observed color, the surface profile of the galaxies and the
color gradient.  With both decreasing surface brightness and color, the true color at
radius $r_n$ is the sum of the intensity weighted colors at $x_n$.  Knowledge of the
surface brightness profile and color gradient allow an iterative procedure to deduce
the true color at $r_n$.
}
\label{beam}
\end{figure}

Some information on this effect can be extracted if the run of surface brightness and
color with radius is known.  For example, one can bootstrap backwards for each
position, iterating on cells of luminosity perpendicular to the surface brightness
profile using the annulus luminosity density as an estimate of the cells in the
pencil beam (see Figure \ref{beam}).  Then the colors from the radial gradient are
assigned to each cell, weighted by cell size and luminosity then summed.  This value
is compared to the integrated color for that radius and iterated until they agree.
The resulting color of the inner cell represents the true internal color at that
radius, in particular, the extrapolation to the galaxy center then represents the
true color at the core.  The calculations depend on the assumed 3D shape that the
pencil beam is passing through. However, numerically, this is only a 5\% correction
for extreme prolate or oblate shapes that is typically much less that the outer
color errors and this correction decreases toward the center.

\begin{figure}[!ht]
\centering
\includegraphics[scale=0.4,angle=0]{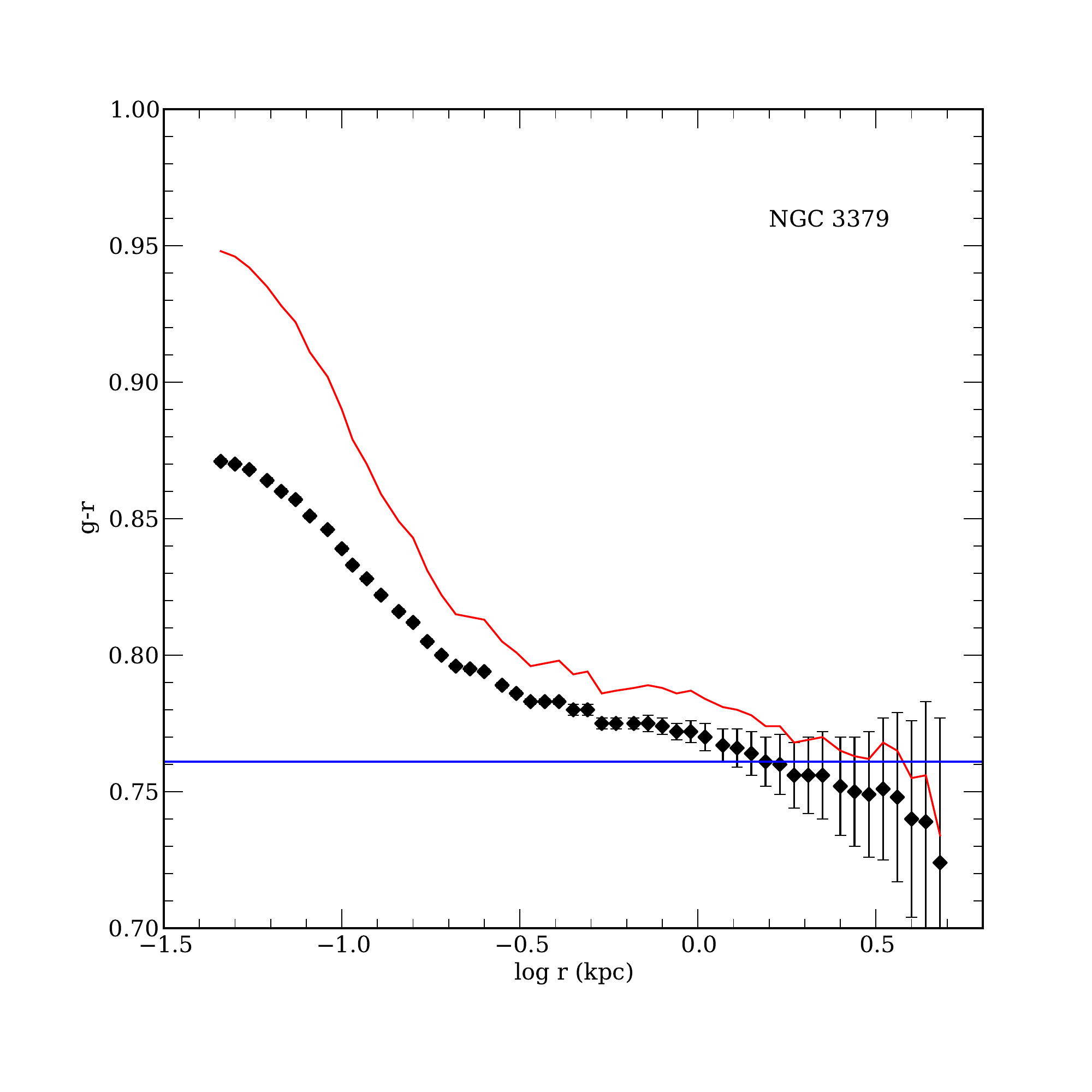}
\caption{\small The run of true color with radius for NGC 3379, an intermediate
luminosity elliptical.  The observed color gradient (in $g-r$) is should as the black
symbols and errorbars.  The deduced true $g-r$ colors as a function of radius are
shown as the red line.  With the typical blue color gradients in ellipticals, the
true color gradient will always be redder than the observed as the line-of-sight
passes through the bluer halo to reach the actual stellar region at radius $r$.  The
mean color of the galaxy is shown as the blue line.
}
\label{N3379}
\end{figure}

An example of this effect on the impact for color gradients is shown in Figure
\ref{N3379}.  The raw color gradient is shown as black data points with the true
internal corrected colors as the red line.  The blue line is the total color of the
galaxy from elliptical apertures.  Note the total color is lower than the mean color
from the surface photometry due to the negative color gradient producing bluer colors
in the outer isophotes that contain more integrated flux than the inner core pixels.
The true color is typically 0.01 to 0.02 redder for intermediate radii, but rises to
0.08 redder in $g-r$ in the central regions.  A true core color will be significantly
redder (or bluer depending on slope of the color gradient) than a total or halo
color.  This will be considered in the discussion section (see S3.3).

\subsection{{\it WISE}}

Differing from the photometry of normal ellipticals in Schombert (2016), we have
added archive data from the {\it WISE} mission to the whole dataset, mostly to offset
the lack of {\it 2MASS} photometry for the dE's in the sample due to the surveys
low limiting magnitude.  The {\it WISE} MIDEX mission (Wright \etal 2010) was a
cryogenically-cooled 40 cm telescope equipped with a camera containing four
mid-infrared focal plane array detectors that simultaneously imaged the same 47x47
arcmin field-of-view.  The entire sky was imaged at 3.4, 4.6, 12 and 22$\mu$m
(labeled as $W1$, $W2$, $W3$ and $W4$ filters) using a HgCdTe 1024x1024 array with
2.76 arcsecs per pixel plate scale for the $W1$ filter.  Processed images, obtained
from IRSA, are flattened, sky-subtracted, calibrated frames with plate scales of
1.375 arcsecs per pixel and resolution of 8.5 arcsec PSF.  The PSF is poor, compared
to our other datasets, but adequate to obtain Holmberg magnitudes and colors.

\begin{figure}[!ht]
\centering
\includegraphics[scale=0.4,angle=0]{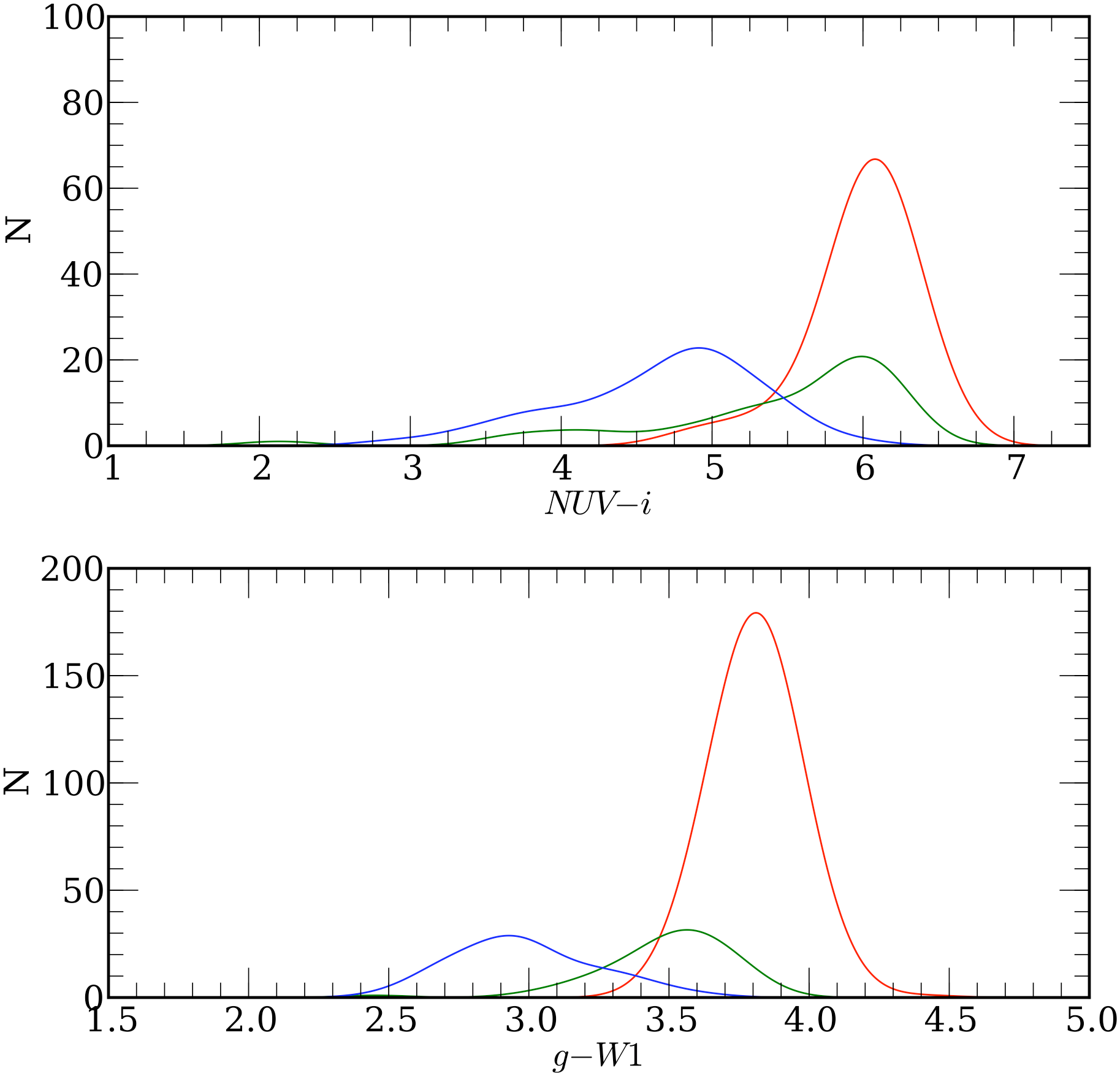}
\caption{\small Normalized histograms for the sample through two colors, $NUV-i$ and
$g-W1$.  The bright elliptical sample (252 galaxies, $M_g < -20$) is shown in red,
the faint elliptical sample (60 galaxies, $-20 > M_g > -16$) in green and the dwarf
ellipticals in blue.  The color-magnitude effect is evident, although the dwarf
sample, selected by morphology, still distinguishes itself by color as well.  Notice,
the range in $g-W1$ color is similar to canonical values in $V-K$, as the two color
map linearly into each other.
}
\label{color_hist}
\end{figure}

The dwarf ellipticals in this sample were all too faint to be detected in {\it 2MASS}
$K$ images, and very few were targeted by {\it Spitzer}.  Thus, comparison to the
bright elliptical sample was problematic in the near-IR.  Instead, the {\it WISE}
archive was searched for the dE's in our samples with GALEX and SDSS coverage (as
well as the bright and faint elliptical samples).  In total, 60 dE's from the DF
catalog met these criteria and have a full range of wavelength coverage.  All the
ellipticals in the bright (252) and faint (60) samples had {\it WISE} coverage.
Figure \ref{color_hist} displays two colors that cover all three major datasets
($GALEX$ $NUV$, $SDSS$ $ugri$ and $WISE$ $W1$) and for the three components of our
sample (bright ellipticals at $M_g < -20$, faint ellipticals between $-20 > M_g >
-16$ and the dwarf ellipticals selected by morphology).  The dispersion is not due to
photometric errors, rather the well-known color-magnitude relation, that is more
evident at long wavelengths than short (Bouquin \etal 2015).  The dwarf ellipticals
notably distinguish themselves in visual appearance, surface brightness profile shape
and color.

It is worth exploring the behavior of the three near-IR filters from {\it 2MASS},
{\it WISE} and {\it Spitzer} ($K$, $W1$ and 3.6$\mu$m) since most SED models produce
$K$ colors and, hopefully, can be easily converted to $W1$ or 3.6$\mu$m.  These filters
have wavelength centers at 2.16$\mu$ ({\it 2MASS} $K_s$), 3.35$\mu$m ({\it WISE}
$W1$) and 3.55$\mu$m ({\it Spitzer} 3.6) respectfully.  The bandwidths are
0.26$\mu$m, 0.66$\mu$m and 0.75$\mu$m, where {\it 2MASS} $K$ is in the Johnson system
and $W1$ and 3.6$\mu$m are in the AB magnitude system.  The sample of ellipticals with $K$,
$W1$ and 3.6$\mu$m observations were culled from the main sample.  There were 253
ellipticals with $K-W1$ colors, 80 ellipticals with $W1-3.6$ colors.  The sample was
taken as a whole and mean colors were deduced using a jackknife average.  Then the
sample was divided into two subsets at the midpoint in luminosity, one brighter than
$M_g = -21$, the other fainter.  The resulting values, plus dispersions are listed in
Table 1.

\begin{deluxetable}{lcccccc}
\tablecolumns{7}
\small
\tablewidth{0pt}
\tablecaption{Mean Near-IR Colors}

\tablehead{
\\
\colhead{$M_g$} &
\colhead{$K-W1$} &
\colhead{N} &
\colhead{$W1-3.6$} &
\colhead{N} &
\colhead{$K-3.6$} &
\colhead{N} \\
}

\startdata

$< -21$ & 0.075$\pm$0.067 & 160 & 0.089$\pm$0.028 & 39 & 0.177$\pm$0.074 & 42 \\
$> -21$ & 0.059$\pm$0.077 &  93 & 0.081$\pm$0.072 & 41 & 0.148$\pm$0.056 & 40 \\
& & & & & & \\
total   & 0.070$\pm$0.073 & 253 & 0.086$\pm$0.029 & 80 & 0.162$\pm$0.066 & 82 \\

\enddata
\end{deluxetable}

The small change in the near-IR colors is unsurprising given that each filter
is close to each other in central wavelength, plus the SED of an old stellar
population peeks around 3$\mu$m and each filter is basically sampling a flat portion
of a galaxy's SED with little dependence on the metallicity of the galaxies.  There
is a slight change in color with absolute luminosity for $K-W1$ and $K-3.6$ (0.016
and 0.019 respectfully), but this change is barely significant and formal fits to the
near-IR CMR produces a correlation coefficient of less than 0.15.

There is also very little evidence of a color term in the two color diagrams where
either $K-W1$, $W1-3.6$ and $K-3.6$ are one axis.  For example, in $u-i$ versus the
near-IR colors there is a range from 2.0 to 3.4 in $u-i$, but mean $K-W1$ color only
varies from 0.055 to 0.070 with a dispersion of 0.085.  Any color variation is minor
compared to the internal errors on the aperture colors.  With respect to correcting
the $K$ magnitudes and colors from SED models to $W1$, we have adopted a $K-W1$ color
of 0.070 with the caveat that for the bluest dwarf ellipticals an addition correction
of 0.010 could be applied and limits the interpretation of age and metallicity by
that amount.

\section{Discussion}

The analysis of the colors of dwarf ellipticals follows the procedure outlined in
Schombert (2016) for bright ellipticals; the empirical two-color relationships for
all the optical and near-IR filters, comparison to multi-metallicity population
models and the color-magnitude relation.  For clarity, we divide the discussion
of the two-color relations into those defined by filters close in wavelength (near
colors) and those whose filters are widely separated in wavelength (far colors).

\subsection{Near Colors}

\begin{figure*}[!ht]
\centering
\includegraphics[scale=0.85,angle=0]{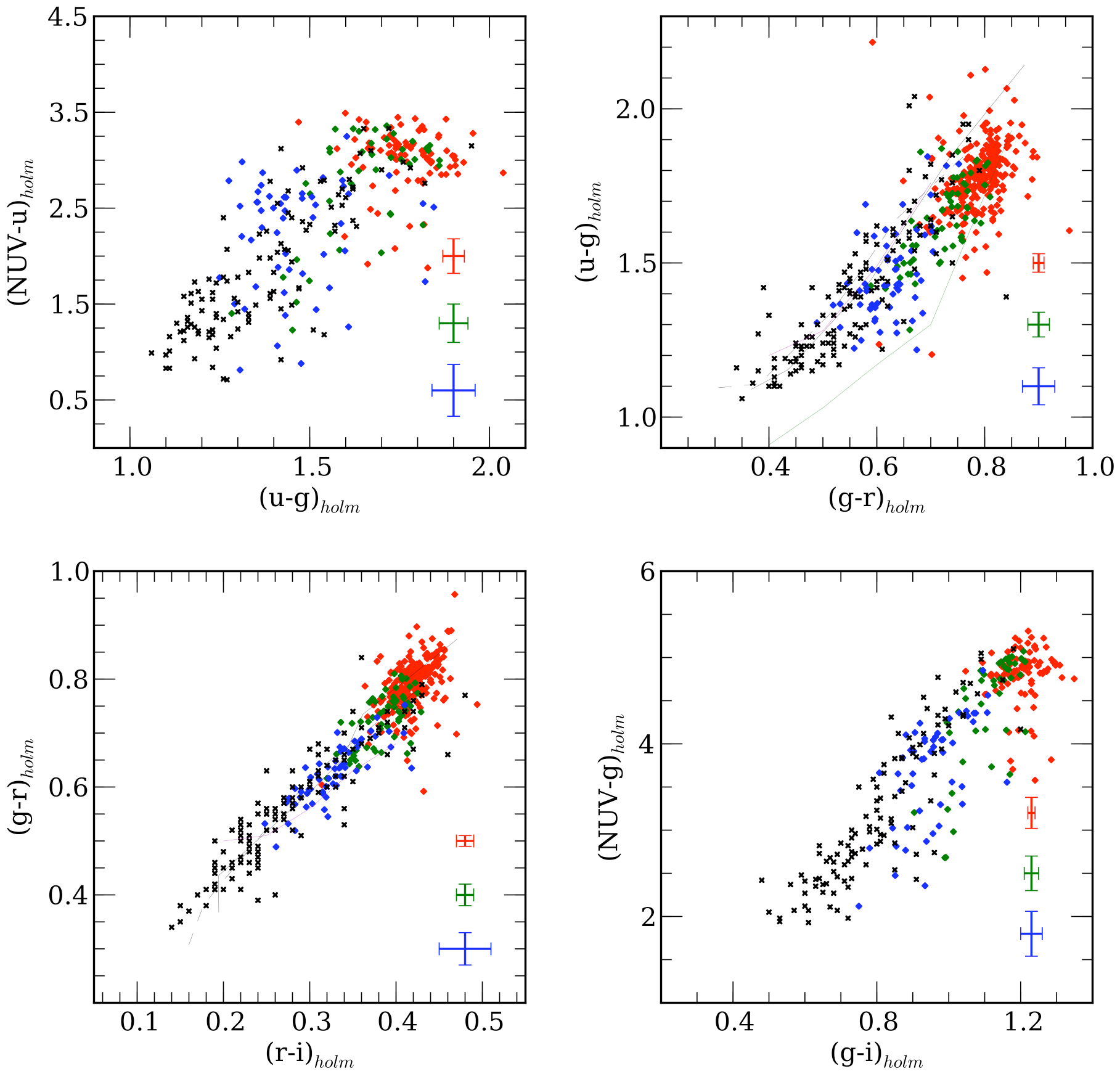}
\caption{\small The two-color relations for $GALEX$ and $SDSS$ near colors
($NUV-u$, $u-g$, $g-r$, $r-i$).  Red symbols are bright ellipticals ($M_g < -20$),
green symbols are faint ellipticals ($M_g > -20$), blue symbols are dwarf ellipticals
(dE) and black crosses are M31 and MW globular clusters.  All colors are based on
Holmberg aperture luminosity values (the Holmberg radius determined in the $g$ frames
and applied to the other filters).  Typical error bars as shown on the right side of
every panel.  The solid black line is a 12 Gyrs SSP model, the dashed line is a 5
Gyrs model (varying from [Fe/H]=$-$2.5 to $+$0.3).  The magenta line is the average
colors 1850 globulars in the Virgo cluster (Powalka \etal 2016).  The green line is the
locus of RC3 Sb-Sm galaxies, i.e. a locus of star-forming colors.
}
\label{4panel_near}
\end{figure*}

The colors of the new faint and dwarf ellipticals in our sample extend the relations
found in Schombert (2016) by six more magnitudes in luminosity.  The near colors are
defined by the near-UV and optical filters ($GALEX$ $NUV$, $SDSS$ $ugri$).  Four of these
two-color relations, between neighbor filters, are shown in Figure \ref{4panel_near}
along with MW and M31 globular cluster colors from Galleti \etal (2009) and Peacock
\etal (2010).  The mean error bars for each of the samples is also shown.  In
general, the colors are coherent from each of the samples, meaning that galaxies that
are blue (or red) in one color set are also blue (or red) in other filter
combinations.  The exception are $NUV$ colors that display a turnover at the reddest
colors.  As discussed in Schombert (2016), this behavior is well modeled by the single
abundance scenario proposed by Yi \etal (1998) and underlies the importance of a
small metal-poor population to colors even in massive ellipticals dominated by
metal-rich stars (see Rakos \etal 2008).

Also shown for the SDSS colors ($u,g,r,i$) are stellar population models for 12 and 5
Gyrs for varying mean metallicities.  Again, as discussed in Schombert (2016), the
bluer colors ($u-g$ vs $g-r$) indicate that an internal metallicity distribution is
required to match the global colors (so-called multi-metallicity models, Schombert \&
Rakos 2009).  Where the bluer $u-g$ colors are influenced by a small metal-poor
population (also seen in the $NUV$ colors), presumably a population of between 5 to
10\% of the total stellar mass that is composed of the first generation of stars with
near globular cluster metallicities.  The multi-metallicity models (shown in
Schombert 2016) follow the bright ellipticals colors to a greater degree than single
metallicity models (SSP, see Schombert 2016).

The SDSS colors display the well-known degeneracy between age and metallicity.
The 12 and 5 Gyr SSP tracks are indistinguishable in two-color space, although the 5
Gyr models require extremely high metallicities ([Fe/H] $>$ 0.6) to match the colors
of the brightest ellipticals.  In a single two-color comparison, it is impossible to
separate the possibility of very young stellar populations with high metallicities
versus older stellar populations with low metallicities.  Separation can be achieved
with a longer wavelength baseline in color due to the fact that age and metallicity,
while still coupled at all optical and near-IR colors, have varying contributions to
color as one goes to longer wavelengths (see \S 3.3).

As noted by Schombert (2016), the trend for positively correlated colors is seen in
all colors except for $NUV-u$.  Elliptical colors are, in general, coherent from
filter to filter (i.e., red galaxies are red in all filters). In addition, outliers
are usually outliers across all their colors with the probable explanation that some
contamination is in the galaxy itself or undetected strong emission lines or simply a
flaw in the calibration and/or reduction pipeline. Their rarity did not warrant extensive
investigation as they are not relevant to the averaged results.  Correlations between
colors have the lowest scatter for widely spaced filters, mostly because filters
close in wavelength have less dynamic range (the slope of the galaxy SED varies
little over small changes in wavelength except near the 4000\AA\ break) and
photometric errors play an increasing role. 

A criterion for recent star formation in bright ellipticals was a $NUV-r$ cutoff at
5.5 (Schawinski \etal 2007) of which 93\% obey this selection.  However, for
luminosities fainter than $-$20, only 40\% had redder colors and, for the dE sample,
only 3\% have redder colors.  This probably does not reflect a steep increase in
recent SF for lower mass ellipticals but, rather, is the effect of decreasing
metallicity where colors bluer than $NUV-r = 5.5$ are typical for old stellar
populations with mean [Fe/H] $< -0.5$ (Schombert 2016).  Younger mean age is not
ruled out in dE's based solely on $NUV$ colors, however, UV colors, by themselves, can not
conclusively demonstrate younger age.  The locus of late-type RC3 galaxies is
shown in the upper right panel of Figure \ref{4panel_near}.  At the blue end of this
sequence are galaxies with current star formation rates between 0.10 and 1 $M_{\sun}$
per yr.  Only a handful of dE's approach this color realm.

With respect to the three ellipticals groups (bright, faint and dwarf), the trend for
bluer colors in all filters is evident.  Where the bright ellipticals defined an
extrapolation of the globular cluster colors in all filters, the faint ellipticals
clearly overlap with the reddest globulars and the dwarf ellipticals are consistent
with the mean globular clusters leaving only the most metal-poor globulars at the extreme
end in color.  If color maps directly into [Fe/H] (which it probably does not for the
dE's, see \S3.4), then the faint ellipticals have [Fe/H] values slightly less than
solar and dE's range from $-$1.5 to $-$0.5 in average metallicity.  This is
consistent with the CMR (see \S3.3).

There is some tendency for the dwarf ellipticals to be slightly bluer, on average,
than the globulars in the bluest colors ($NUV-g$ and $u-g$).  However, they are in
agreement for the redder colors ($g-r$ and $r-i$).  Any deduction of [Fe/H] from
colors, as in Schombert (2016), assumes a 12 Gyrs age and agreement across the
optical and near-IR colors with the bright ellipticals whose metallicity calibration is
set by the globular cluster two-color relations.  We will explore this discrepancy as
it impacts their estimate of mean age and metallicity for dE's in \S3.4.

We also note the recently published SDSS and $K$ colors for a broad sample of Virgo
globular clusters (Powalka \etal 2016) is shown in Figure \ref{4panel_near} as a
magenta line.  This line represents a moving average in color space of 1850
globulars.  The trend in two-color space is identical to MW and M31 globulars,
although there is a tendency to slightly redder blue colors at the high metallicity
end of the Virgo globulars sequence.  Whether this signals a break in the
age-metallicity relation for globulars in Virgo, compared to the Milky Way and M31,
or a shift in the color calibration at redder colors is unclear.  We will continue to
use the MW/M31 sequence to define the 12 Gyrs locus and calibration of [Fe/H] for
ellipticals as outlined in Schombert (2016).

\subsection{Far Colors}

\begin{figure*}[!ht]
\centering
\includegraphics[scale=0.85,angle=0]{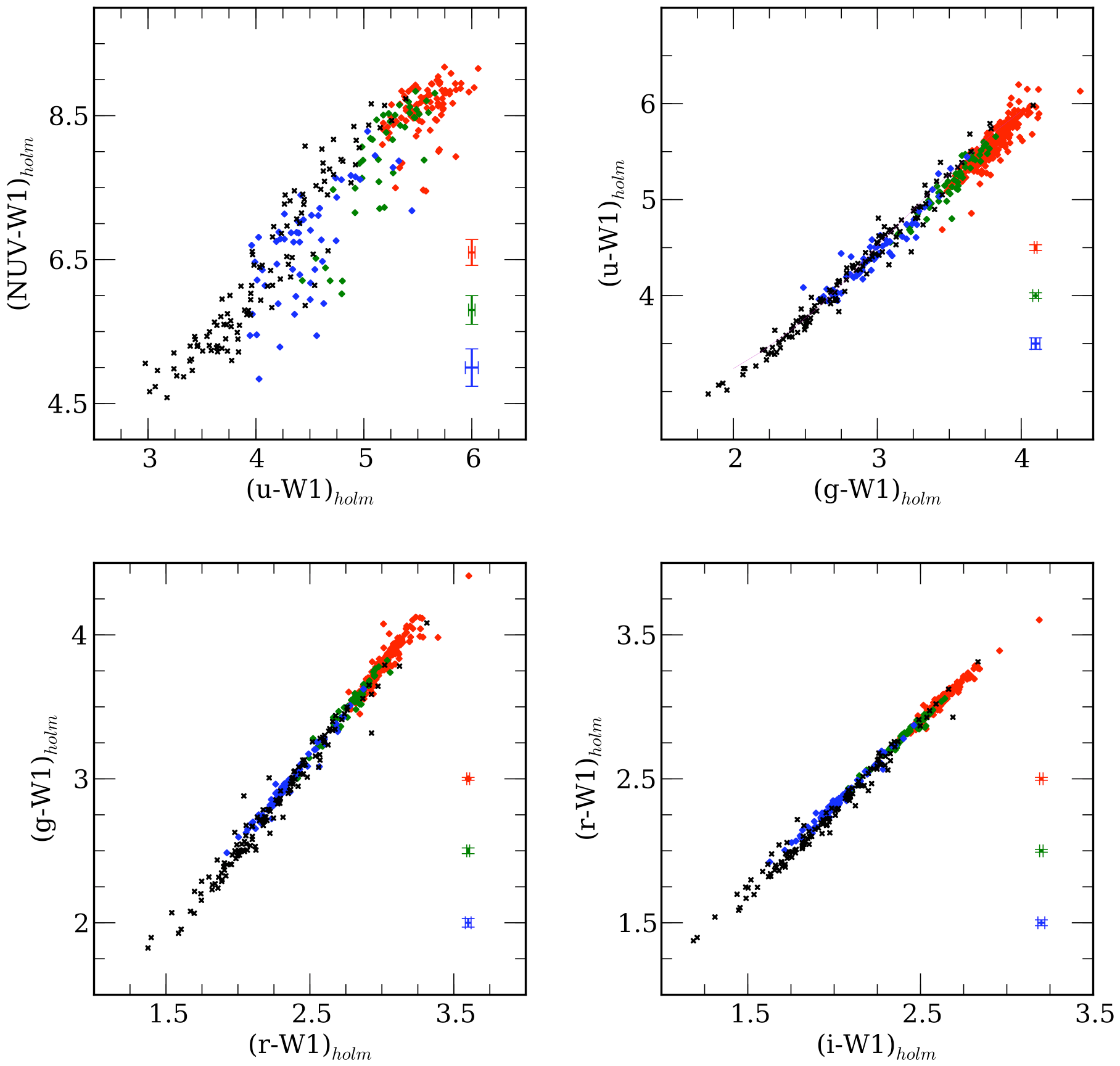}
\caption{\small The two-color relations for $GALEX$, $SDSS$ and $WISE$ far colors
($NUV-W1$, $u-W1$, $g-W1$, $r-W1$, $i-W1$).  Red symbols are bright ellipticals ($M_g < -20$),
green symbols are faint ellipticals ($M_g > -20$), blue symbols are dwarf ellipticals
(dE) and black crosses are M31 and MW globular clusters.  All colors are based on
Holmberg aperture luminosity values (the Holmberg radius determined in the $g$ frames
and applied to the other filters).  Typical error bars as shown on the right side of
every panel.  The solid black line is a 12 Gyrs SSP model, the dashed line is a 5
Gyrs model (varying from [Fe/H]=$-$2.5 to $+$0.3).  The magenta line is the average
colors 1850 globulars in the Virgo cluster (Powalka \etal 2016).
}
\label{4panel_far}
\end{figure*}

The similarity between normal and dwarf ellipticals, and globular clusters, continues
in the colors with the largest wavelength separation shown in Figure
\ref{4panel_far}.  Here, the $WISE$ filter $W1$ replaces the $2MASS$ $K$ filter from
Schombert (2016) with only a nominal correction to stellar population models to jump
from $K$ colors to $W1$ colors.  Again, the trends from Schombert (2016) are
reproduced, now with a greater range of galaxy luminosities (i.e. bluer colors).  The
optical to near-IR colors display high uniformity from bright to dwarf ellipticals.
This would rule out a strong AGB contribution for dwarf ellipticals compared to
normal ellipticals as this would be signaled in a sharp change in slope in optical to
near-IR two-color diagrams (see Schombert \& McGaugh 2014).  A lack of AGB colors, in
turn, rules out an strong star formation in the last few Gyrs.

As with the near colors in Figure \ref{4panel_near}, the far colors also display the
behavior of the dwarf ellipticals overlapping with the intermediate metallicity
globular clusters ([Fe/H] between $-$1.5 and $-$0.5).  SSP models for the far colors
have very little separation by age, tracking parallel to color-color sequences with
only metallicity to provide the sole variation in color.  Thus, these individual
colors, by themselves do not distinguish between a subsolar dE population of 12 Gyrs
or a near solar metallicity dE population, but with ages less than a Gyr.

However, there is a subtle change as one goes from $u-W1$ to $i-W1$.  While the
bright and faint normal ellipticals maintain their relative distributions with
respect to globular cluster colors, the dwarf ellipticals become increasing redder
with respect to the mean globular color.  This can been seen by using the mean color
of the globular clusters and bright ellipticals as an anchor to the reddest and
bluest colors from $u$ to $W1$.  The mean color of the faint elliptical sample lies
at 0.71, 0.69, 0.69 and 0.68 in the fraction of this interval for $u-W1$, $g-W1$,
$r-W1$ and $i-W1$.  On the other hand, the mean dE sample color lies at 0.26, 0.19,
0.16 and 0.12 for the same colors and interval.  In other words, the mean color for
dE's decreases, with respect to normal ellipticals, as we go from $u-W1$ to $i-W1$.

This behavior is unexpected since the color of normal ellipticals with respect to the
globular clusters was extremely consistent from $GALEX$ to $Spitzer$ (Schombert
2016).  In fact, assigning a metallicity value, just from a normal ellipticals color
using globulars as a calibration, produced a consistent and robust measure even if
the scatter in an individual color was high.  This procedure will fail for dwarf
ellipticals as using colors closer in wavelength will result in increasing low
[Fe/H] values.  The most obvious explanation for this type of color behavior is an
age effect, that was first indicated with narrow band colors (Rakos \& Schombert
2004) and will be discussed in \S3.4.

\subsection{Color-Magnitude Relation}

It was demonstrated in Schombert (2016) that the CMR is primarily a relationship
between stellar luminosity (a proxy for total galaxy stellar mass) and mean
metallicity.  While limited age and recent star formation effects can not be
completely excluded (e.g., Faber \etal 1995), these effects are at the 5\% level for
colors (although may have a larger contribution for spectral indices studies).  The
slopes for the CMR, found in Table 3 of Schombert (2016), are identical to slopes
from other CMR studies (Bernardi \etal 2003, Chang \etal 2006).  Changes in the slope
of the CMR with redshift is an interesting measure of galaxy chemical evolution, so
the accurate slope values for zero redshift samples is an important parameter.

\begin{figure}[!ht]
\centering
\includegraphics[scale=0.5,angle=0]{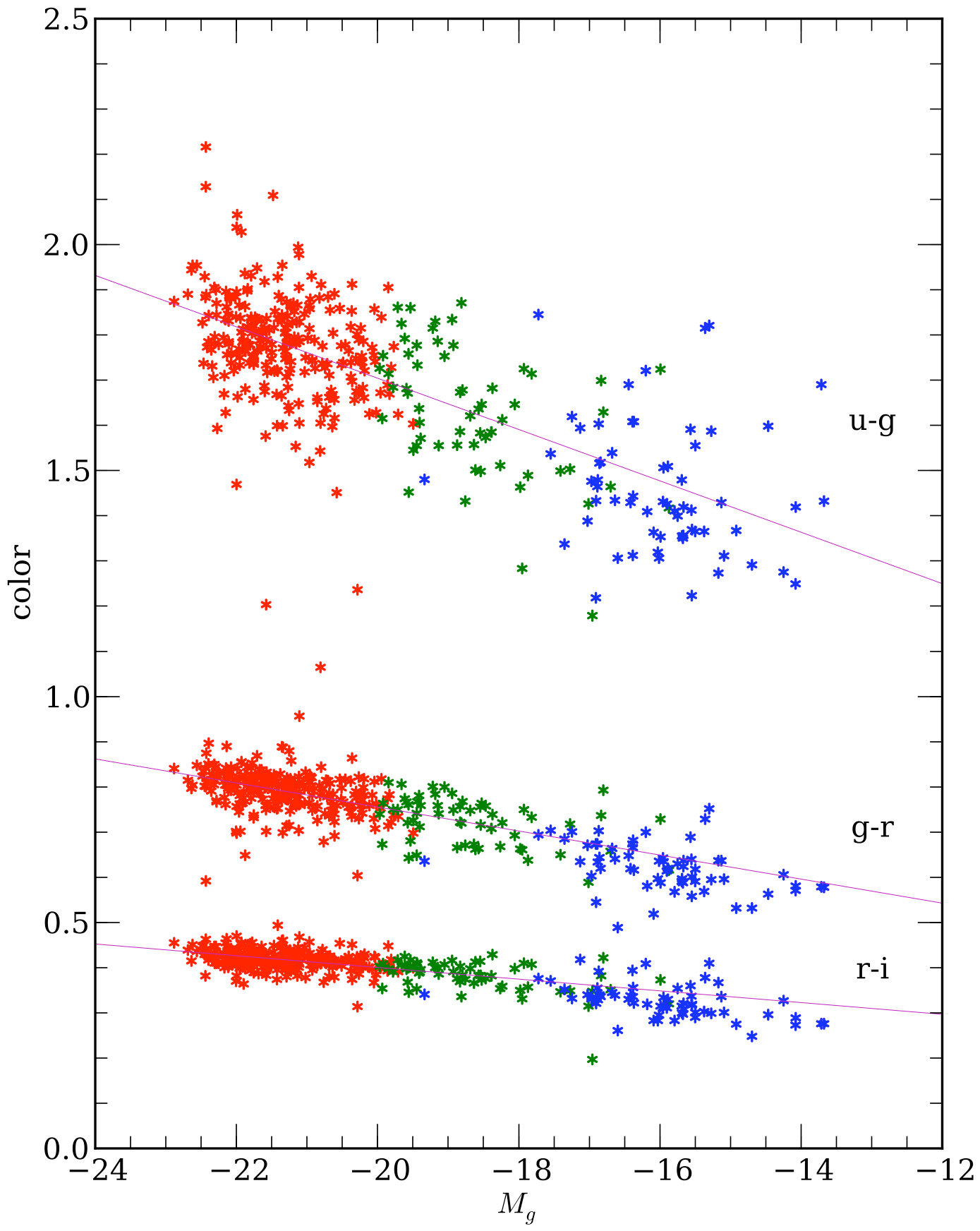}
\caption{\small The CMR for near optical colors.  Red symbols are bright ellipticals
$M_g < -20$) with power-law shaped profiles.  Green symbols are faint ellipticals
($M_g > -20$) with power-law shaped profiles.  Blue symbols are dwarf ellipticals by
morphology.  The bright and faint ellipticals follow the same linear fit (slope and
zeropoint).  While the dE's are slightly bluer in $g-r$ and $r-i$ from the normal
elliptical fit (e.g., 80\% of the dE's are below the fit in $r-i$.  The deviation
becomes more prominent at redder colors.
}
\label{cmr_near}
\end{figure}

\begin{figure}[!ht]
\centering
\includegraphics[scale=0.5,angle=0]{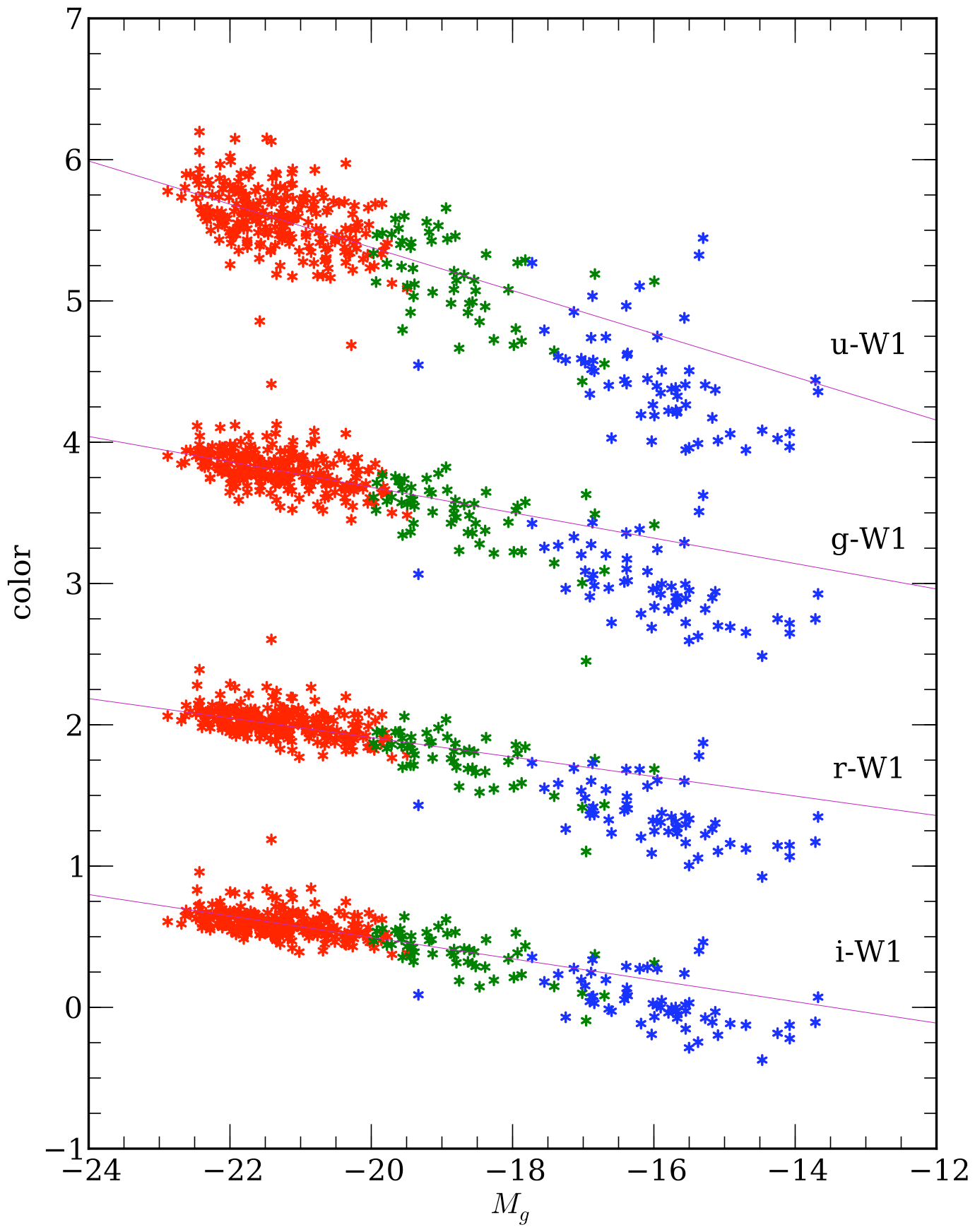}
\caption{\small The CMR for far color, optical to near-IR.  Symbol colors are the
same as Figure \ref{cmr_near}.  The deviation of dwarf ellipticals from the normal
elliptical CMR is more obvious in the near-IR colors.  The sequence from bright to
fainter dE's deviates farther from the normal elliptical fit with decreasing
luminosity.  However, even a majority of brightest dwarf ellipticals are below the
normal ellipticals CMR.
}
\label{cmr_W1}
\end{figure}

The CMR for near and far colors are shown in Figures \ref{cmr_near} and \ref{cmr_W1}.
There is no evidence that the new sample of faint ellipticals deviates from the CMR
slopes fit to the bright ellipticals (shown in each Figure as green symbols).  This
makes a solid statement that {\it power-law shaped ellipticals, that define the
bright and faint ellipticals sample (Schombert 2017), also obey the same structural
and stellar population relationships}.  This will be somewhat of a challenge to many
hierarchical models of galaxy formation which predict extended epochs of star
formation that varies significantly with mass (Naab \etal 2007).  However, age
determination with colors allows for a great deal of flexibility with respect to the
first epoch of star formation and its initial duration (for example, a shorter
duration at a later epoch will mimic a long duration SF event at early epochs, see below).

There is no evidence in any color combination that normal ellipticals are composed of
a significant stellar population with ages less than 12 Gyrs (Rakos, Odell \&
Schombert 2008, Schombert 2016, although see a dissenting view in Graves, Faber \&
Schiavon 2010).  And, thus, the slopes of the various CMR's are consistent, across
all the wavelengths, with a pure metallicity interpretation in a generally old
stellar population.  Even small variations in mean age (greater than 4 Gyrs) would
result in different slopes between blue and near-IR colors (if age varied with
stellar mass, see Eigenthaler \& Zeilinger 2013; Smith \etal 2012).  As demonstrated
in Schombert (2016), there is no indication of an age effect in the CMR, and the new
sample of fainter ellipticals supports this conclusion.  This implies that the CMR
is, in fact, purely a metallicity relationship, presumingly between the stellar mass
that produces metals and, later, produces the galactic winds that remove the
remaining gas and halt star formation and chemical enrichment (see Matteucci 2007).  

The identical CMR slopes for bright and faint ellipticals implies that we can use the
same techniques of assigning a mean metallicity to the new, low luminosity sample as
we did the brighter ellipticals in Schombert (2016).  As outlined in Schombert
(2016), each color can be assigned a metallicity-color relation as guided by the
predictions from multi-metallicity population models (Schombert \& Rakos 2009) tied
to the globular cluster [Fe/H] values.  A guide to the accuracy of this method is to
compare the spread in [Fe/H] as deduced from colors.  This technique produced a mean
dispersion of 0.15 dex in metallicity for both the bright and faint elliptical
samples.  Mapped onto the CMR, this results in [Fe/H] values of 0.5 for the brightest
ellipticals decreasing to $-$0.2 for ellipticals around $-$19.

The behavior of the colors for dwarf ellipticals differs from the normal ellipticals
in that a majority (varying between 60 and 80\%) have colors bluer than expected for
their luminosities based on a linear extrapolation of the CMR.  Small number
statistics make it impossible to determine if the dwarf ellipticals form their own
sequence or are a non-linear extension to the normal elliptical sequence.  We can
only state that the brightest dwarf ellipticals have colors near the CMR and the
deviation becomes greater at lower luminosities.  And, while most the dE's in the
sample have colors that place them below an extrapolation of the CMR from normal
ellipticals, that deviation is greatest at the longest wavelengths.  From the near-IR
colors there is a clear indication that the dwarf ellipticals deviate from the
normal elliptical to a greater extent with decreasing luminosity than is seen in the
optical colors.  

If one assumes that metallicity is still the primary driver for the CMR in dwarf
ellipticals and one applies the same metallicity-color relations for normal
ellipticals to dwarf ellipticals, then the CMR for dE's implies [Fe/H] values near
$-$0.2 for the brightest dE's decreasing to $-$1.0 for the faintest dE's in our
sample.  However, this is not consistent between the colors.  For example, the $u$
and $W1$ colors derive [Fe/H] values near $-$1.5 for the faintest dE's while the
$gri$ colors derive much higher values near $-$0.5 for the same galaxies.  This type
of behavior, not found in normal ellipticals suggests another parameter, the most
obvious being age is in play (although one could entertain extreme IMF and extinction
values).

\subsection{Younger Age for Dwarf Ellipticals}

It is possible to estimate the magnitude of an age effect on the complete sample of
dwarf ellipticals, without assigning a specific age to each galaxy.  The technique
follows the prescription outlined in Schombert (2016) with respect to bright
ellipticals.  A metallicity can be assigned to each color combination based on a
procedure of taking SSP models and convolving them to a multi-metallicity framework
(one that assumes an underlying metallicity distribution based on high resolution
studies of nearby ellipticals, see Monachesi \etal 2011).  The zeropoint to these
multi-metallicity models is set by globular cluster colors (both MW and M31 samples)
whose mean [Fe/H] values are determined directly from their CMD's.  As discussed in
Schombert (2016), single population models are a poor description of bright
elliptical colors, whereas multi-metallicity predictions reproduce the mean colors
and CMR of bright ellipticals without any need to introduce a younger component to
a pure 12 Gyrs population.

\begin{figure*}[!ht]
\centering
\includegraphics[scale=0.8,angle=0]{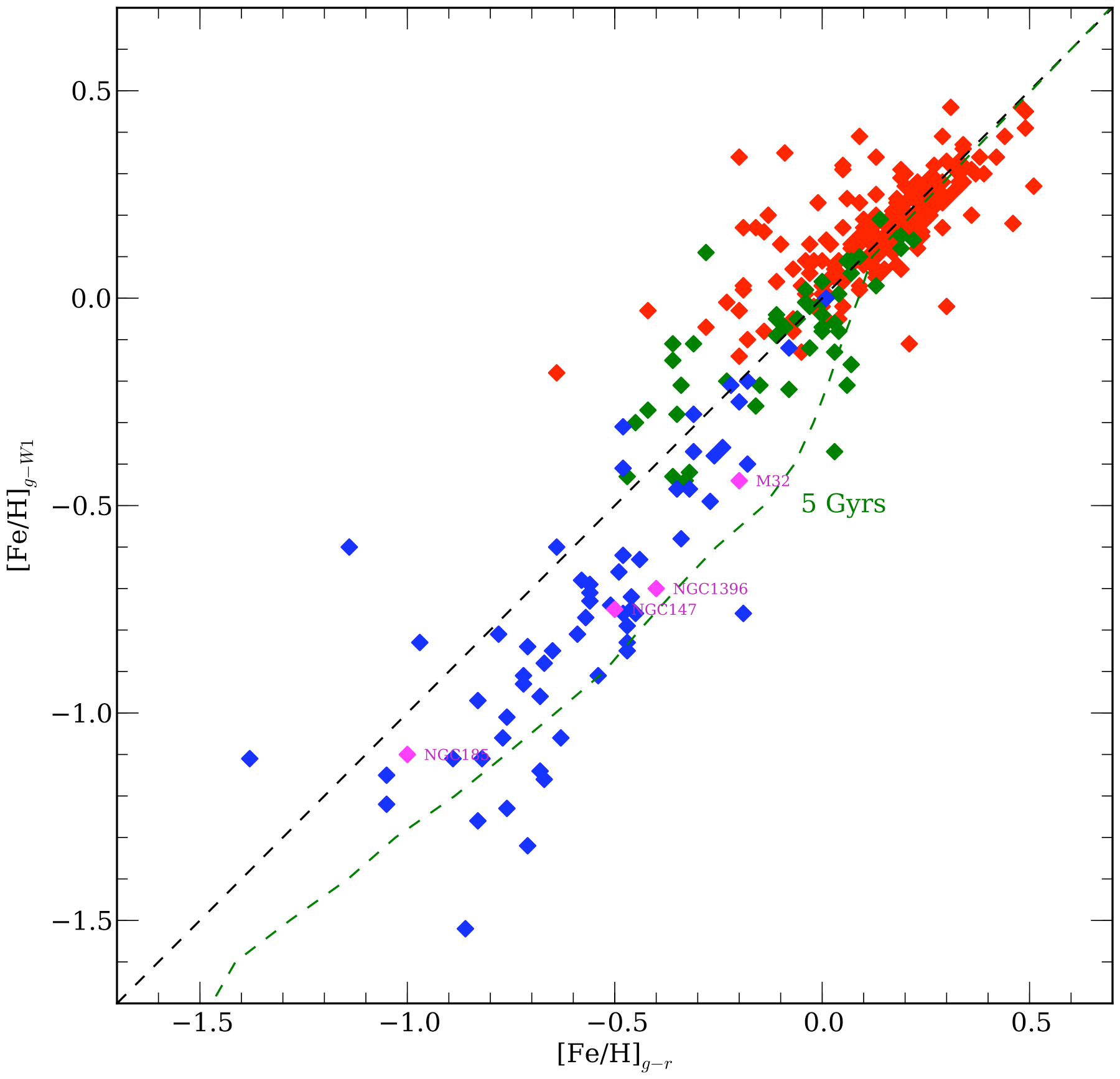}
\caption{\small An example of [Fe/H] deduced from colors, calibrated to the MW/M31
globular cluster metallicities (Schombert 2016).  Bright and faint ellipticals
produce coherent [Fe/H] values from color to color ($\sigma = +0.15$ dex). However,
the dwarf elliptical sample deviates from the unity line is a systematic fashion with
color that indicates an age effect.  A 5 Gyrs model is placed in the same color-color
space which deviates downward for metallicities deduced from near-IR colors.
dE's range from 2 to 8 Gyrs younger than normal ellipticals, with mean age decreasing with
decreasing dwarf mass.  Also the positions 
for three dE's in the Local Group (M32, NGC147 and NGC185) plus one Fornax dE
(NGC1396) are indicated using their CMD determined age and metallicities with respect
to the models.
}
\label{age_feh}
\end{figure*}

Using these color to metallicity models, and assuming a 12 Gyrs age, each color
predicts a [Fe/H] value.  Any systematic deviations from a mean [Fe/H] based on an
average of all the colors would signal a problem for the technique, such as a 2nd
parameter other than metallicity (e.g., age or IMF variations) as the main
determinant in the continuum shape of galaxy spectra (i.e., colors).  None were seen
in the faint plus bright elliptical samples and the scatter in the deduced [Fe/H]
values were solely owing to error in the colors.

We can now apply those metallicity calibrations to the dwarf elliptical colors, again
with the assumption of a solely metallicity driven continuum and a mean age of 12
Gyrs.  As expected from the deviations in the CMR, this procedure fails.  The colors
for faint ellipticals are in line with the bright ellipticals (see Figure
\ref{age_feh}), and even the brightest dE's are consistent from color to color.  But
a majority of the dE's do not produce consistent [Fe/H] values.  Typically, they
overestimate [Fe/H] for blue colors and underestimate [Fe/H] for red and near-IR
colors.  One such color combination is shown in Figure \ref{age_feh} for the
metallicity deduced from $g-r$ versus the metallicity deduced from $g-W1$.  The
normal ellipticals display a good one-to-one correspondence between the calculated
[Fe/H] values.  However, the dwarf ellipticals begin to deviate below metallicities
of $-$0.3 and a majority have near-IR calculated [Fe/H] values 0.5 dex below those
calculated from optical colors.

If one relaxes the age requirement of 12 Gyrs, for example assuming a mean age of 5
Gyrs, then the one gets the resulting metallicity track as shown in Figure \ref{age_feh} 
(green dotted track).  The deviation in color is well matched to an age effect, but now the
deduced [Fe/H] values become non-unique as there is a wide range of possible age and
metallicity combinations that produce the observed colors.  Some of this range can be
narrowed by using multi-age tracks across several color combinations.  The result of
this numerical experiment is such that dwarf ellipticals seem to range in age from 12
to 6 Gyrs in age, and $-$0.3 and $-$1.4 in [Fe/H].  There is a clear trend of
decreasing metallicity and decreasing age with luminosity (stellar mass).  The
positions of the Local Group and Fornax dE's, with CMD determined ages and
metallicities, are indicated using the model predictions.  Their positions agree well
with the general trend of the dE sample.

\section{Conclusions}

A difference between the CMR for normal and dwarf ellipticals can be traced back to
Caldwell (1983).  His Figure 3 displays the $U-V$ colors for a small sample of Virgo
bright ellipticals plus a sample of 19 dE's.  While the bright end of the CMR is
ill-defined, the dwarf ellipticals clearly lie blueward of a linear extrapolation of
the bright elliptical trend.  Interpretation at that time focused on recent star
formation in dwarf ellipticals, but concluded that massive stars were missing and the
bluer colors were due to a younger mean age.

Some curvature in the optical CMR, at high and low luminosities, was detected by
Bernardi \etal (2011).  They covered a range of $-$18 to $-$24 in SDSS $r$, compared
to our study from $-$15 to $-$24.  We do not, statistically, detect an upturn at high
luminosities in our sample, but the amount they claim is within our errors.  An
upward turn at high luminosities is interpreted by Bernardi \etal as a signal of
major mergers in the history of bright ellipticals.  Our data does not confirm or
falsify this conclusion.  The shape of the CMR at the bright end appears to be
independent of environment (De Propris, Phillip \& Bremer 2013), supporting our old
age, pure metallicity interpretation as the internal process of chemical enrichment
dominates.  The downturn they find at low luminosities is at the same luminosity that
we find in our dwarf sample (first seen by Janz \& Lisker 2009), thus they agree with
our general observations of the low mass end.  Another example of a blueward trend in
the CMR at low luminosities is found in Agulli \etal (2016), who detect a blueward
downturn for red sequence cluster members of Hercules (A2151).  While they can not
separate out galaxies by morphological, there is a clear trend for the red sequence
members of their spectroscopic survey to display bluer colors at luminosities fainter
than $-$18 with respect to the expectation from a linear fit to the bright end of the
CMR (see their Figure 1).

In contrast is the work of Roediger \etal (2016) who studied the CMR in Virgo down to
$M_g = -9$.  They do not identify a downward trend for dE's, but instead find a
surprising flattening of color for galaxies fainter than $-$14, well below our sample
limit.  It is difficult to make a comparison with our sample for their sample numbers
are very low for ellipticals brighter than $-$18 (i.e., the normal elliptical CMR is
ill-defined for their sample), and we have no ellipticals fainter than $-$14 in our
sample.  Their conclusions, that very low mass galaxies quench in star formation at
the same epoch to produce similar colors, does not seem to apply to our dwarf
ellipticals.  A similar problem is found for a test of the linearity of the CMR in
Virgo by Smith Castelli \etal (2013).  In that study the focus was on the $g-z$ CMR
of Virgo ellipticals between $-$16 and $-$20.  Again, the CMR is ill-defined with few
bright ellipticals to stabilize the fitting of the high luminosity end.  However,
there was no evidence of a blueward trend between ellipticals brighter than $-$18 and
those fainter than this demarcation.

Regardless of the filter combination, we confirm the main conclusion first expressed
by Janz \& Lisker (2009) that normal and dwarf ellipticals do not follow one, linear
CMR.  With the assumption that the deviation from normal elliptical CMR for dwarfs is
real, and is systematic across the colors from $GALEX$ to $WISE$, we present an
interpretation based on a younger age for low mass ellipticals.  We base this
conclusion on the fact that the dE's deviate from the metallicity-color relations,
defined by normal ellipticals, in a coherent fashion for each color combination.  A
trend that is predicted by SSP models, calibrated to globular cluster ages and
metallicities, and indicates that the mean age of the stellar population in dwarf
ellipticals is younger than normal ellipticals by between 2 to 8 Gyrs (younger with
decreasing luminosity).  With these younger ages, deduced metallicities range from
$-$0.5 on the high mass side to $-$1.5 for the lower mass dwarfs, in agreement with
the metallicities deduced from dwarf elliptical RGB CMD's (Caldwell 2006).

A younger mean age for dwarf ellipticals comes as no surprise given the detailed SFH
results for M32, our nearest dE (Monachesi \etal 2011).  To summarize the CMD results
for M32, they find a mean metallicity of the RGB of $-$0.2, but with a wide spread
ranging from $-$1.0 to solar.  Analysis of the red clump gives a mean age of 8 to 10
Gyrs (i.e., 2 Gyrs younger than normal ellipticals), but the detection of AGB stars
above the RGB indicates some component of a 5 Gyr intermediate age population.  Using
a mass-weighted analysis, Monachesi \etal (2012) find 40\% of the stellar population
in M32 has an age between 2 and 5 Gyrs, with the remaining stars being older than 5
Gyrs for a average age of 7 Gyrs for all the stars.  For the mean metallicity, this
age agrees well with the color trend in Figure \ref{age_feh} where the dE's with
[Fe/H] values between $-$0.5 and solar have only slightly younger ages than normal
ellipticals (in the 8 Gyrs range).

In a similar analysis, Mentz \etal (2016) find the age and metallicity of NGC1396 (a
well-studied Fornax dE) to be $-$0.4 and 6 Gyrs.  Geha \etal (2015) find values of
$-$1.0 and 10 Gyrs for NGC185 and $-$0.5 and 6 Gyrs for NGC147, the other two dE
companions to M31.  These values, and M32, are shown in Figure \ref{age_feh} and are
consistent with the trend of luminosity and age for the entire dE sample.  In
addition, Rakos \& Schombert (2004) found dE's are 3-4 Gyrs younger than bright
ellipticals in Coma using narrow band continuum colors between 3500\AA\ and 5500\AA.
Thus, a different age for stellar populations in dwarf ellipticals, compared to
bright ellipticals, is well established.

Lastly, Sybilska \etal (2017) find, using SAURON results for 12 Virgo dE's, that
their sample galaxies contain two stellar populations; an old (12 Gyrs) population
and a younger (age $<$ 5 Gyrs) population with varying degrees of dominance.  They
conclude they two populations are due to either extended SF (longer duration) or a
second burst, leaning toward an extended SF interpretation.  Range of ages presented
by Sybilska \etal agrees well with our colors, although the [Fe/H] values are much
lower than we deduce from our multi-metallicity models.  

It is important to note that assigning a mean age by color to dwarf ellipticals does
not distinguish between a scenario where the initial epoch of star formation is at
lower redshifts than normal ellipticals, or if that initial epoch was extending in
duration such that the distribution of stellar ages peaks at the mean age value.  In
fact, the proposal by Thomas \etal (2005) based on spectral indices (particularly the
$\alpha$/Fe ratio), is that lower mass normal ellipticals have increasing SF duration
times.  Here the scenario is that all ellipticals have a common (and early) initial
stage of SF, but lower mass normal ellipticals peak in star formation 
at later epochs (typically 2 to 3 Gyrs later than the most massive
ellipticals).  Given the information gleaned from a handful of nearby dwarf
ellipticals that have high resolution CMD imaging, it appears that later scenario is
more probable as a majority of these nearby dwarfs have a significant fraction of
their stars in resolved old populations with ages greater than 10 Gyrs.  Duration
versus a later epoch of SF can be tested by color models; however, the $\alpha$/Fe
index is a more sensitive indicator of duration, at least for the first few Gyrs
before Type I SN begin to enrich a galaxy with Fe.

In addition, there is no strong motivation to presume the SFH of dwarf ellipticals
resembles normal ellipticals.  Aside from their lack of visible structure (i.e.,
spiral arms) and no SF features (i.e., HII regions), the family of dE's has little
else in common with normal ellipticals as they are structurally and kinematically
distinct from the normal elliptical sequence (Schombert 2017).  It also seems
difficult to understand how normal ellipticals can be built from small objects like
dE's or dIrr's given the differences in their stellar populations, unless this
construction occurs at redshifts before the initial SF epoch.  The fact that the mean
age of dwarf ellipticals appears to approach the mean age of normal ellipticals at
higher luminosities, in a smooth and uniform fashion (see Figure \ref{age_feh})
suggests an independent process guides the SFH of dwarf ellipticals separate from the
SFH process of ellipticals, but ends with similar patterns of SF at the highest dwarf
masses.

The most commonly accepted evolutionary scenario for all types of galaxies since
$z=1$ is the so-called downsizing scenario (Gavazzi \etal 1996; Cowie \etal
1996), with massive galaxies forming the bulk of their stars in short,
high star formation rates (SFR) periods at earlier epochs, whereas less-massive galaxies
have delayed star formation histories which are extended over a longer time period
(Gavazzi \etal 1996; Thomas \etal 2005; Nelan \etal 2005; Jimenez \etal 2007;
Fontanot \etal 2009).  In that limited context, the colors of dwarf ellipticals
presented herein agree with that scenario.  However, there is no evidence that normal
ellipticals display any variation in age (although it is nearly impossible to
distinguish an age change of less than 2 to 3 Gyrs from an old 12 Gyrs population).
Recchi, Calura \& Kroupa (2012) find the $\alpha$/Fe index varies uniformly with
stellar mass and also interpret this relationship as smaller galaxies forming over
longer timescales (downsizing), allowing a larger amount of Fe (mostly produced by
SN Ia) to be released and incorporated into newly forming
stars.  In addition, Fontana \etal (2004) find the typical $M/L$ ratio of massive
ellipticals is larger than that of less massive ones, suggesting that their stellar
population formed at higher redshifts than dwarf ellipticals (see also Cappellari
\etal 2013).

The mechanisms for halting star formation typical falls into two categories;
quenching (Bundy \etal 2006) and exhaustion (S{\'a}nchez-Janssen \etal 2013).  For normal
ellipticals, who appear to have a majority of their stars in the form of an old,
metal-rich population, their $\alpha$/Fe ratios argues that an exhaustion process
dominates.  Presumingly where the gas exhaustion is triggered by removal of large
quantities by galactic winds (Matteucci 2004).  For dwarf ellipticals, since their
existence is strongly tied to a cluster environment (there are very few dE's in the
field, Sandage, Binggeli \& Tammann 1985), a quenching mechanism due to ram pressure stripping is
implied.  If the trend from Figure \ref{age_feh} is generic, then lower mass dwarf
ellipticals have longer durations of initial SF and, therefore, younger mean ages
than higher mass dwarf ellipticals, whose age and [Fe/H] values continue the pure
metallicity sequence of normal ellipticals.  The age difference detected in dE colors
is similar to the trend of age/metallicity seen in Peng, Maiolino \& Cochrane (2015)
study of SDSS galaxies.

The dilemma for a longer duration of SF for lower mass dE's is that longer timescales
should produce higher metallicities, where the opposite is seen, the lowest mass
dwarf ellipticals have the lowest metallicities.  This presumes a 
SFR that is similar by mass.  We can use the stellar mass-SFR correlation for LSB
dwarfs from the SPARC dataset (Lelli, McGaugh \& Schombert 2016) as a crude indicator of
initial SFR.  The so-called main sequence for LSB galaxies has a slope of unity with
a typical SFR of 0.0003 $M_{\odot}$ per yr$^{-1}$ for stellar masses of $10^7$
$M_{\odot}$ rising to 0.04 at a stellar mass of $10^9$ $M_{\odot}$ (McGaugh,
Schombert \& Lelli 2017).  Over durations of 8 to 2 Gyrs with increasing dE mass,
this corresponds to between 10 and 25\% the needed mass to make a present-day dwarf
elliptical.  Thus, the initial star formation rates would need to be only increased
by a factor of 5 at early epochs to produce the objects we see today as quiescent
dwarfs ellipticals.  These low rates of star formation would inhibit chemical enrichment, and the
deduced [Fe/H] values for the dE's in this sample are only slightly metal richer than
present-day LSB dwarfs of similar stellar mass (Schombert \& McGaugh 2015).

\begin{figure*}[!ht]
\centering
\includegraphics[scale=0.8,angle=0]{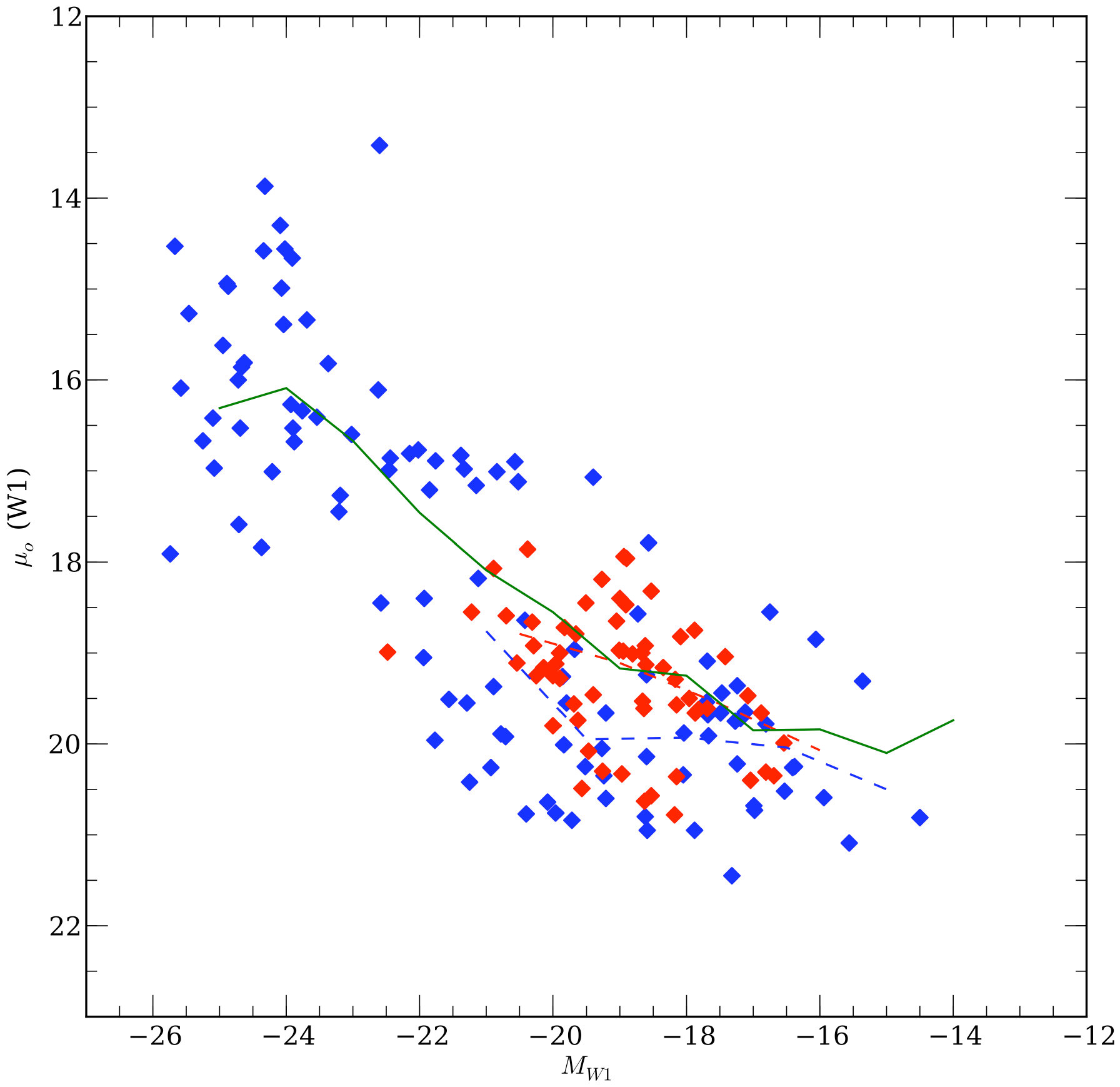}
\caption{\small The relationship between total luminosity (i.e., total stellar mass
at W1) and central surface brightness ($\mu_o$) based on exponential fits to the W1
surface brightness profiles.  The blue symbols are gas-rich LSB dwarfs from the SPARC
database.  Red symbols are the dwarf ellipticals from this study.  The green line is
a moving average of the 11HUGS and James \etal dataset for 462 late-type galaxies.
The dashed lines are the moving average for the SPARC and dE samples.
The dE's overlap the high SFR Irr sample of 11HUGS, and lie slightly higher in
density than present-day LSB galaxies.  The interpretation is that dE's had slightly
higher SFR than present-day LSB's, resulting in slightly higher central stellar
densities.  However, the SF history of dE's would have been remarkably similar to
present-day LSB's, i.e. slow, inhibited SF, until the cluster environment quenched
the SF and removed the remaining gas.  The low stellar densities and similar [Fe/H]
values to present-day LSB's suggests a common star formation history.
}
\label{mu}
\end{figure*}

The last remaining question is what kind of galaxy, in terms of structure, would be
produced from a history of early, but low level SF, which is halted suddenly by
infall into the cluster environment?  The presumption here is that LSB dwarfs and
young dE's follow the same style of star formation at similar rates during the era of
galaxy formation.  The gas is converted into stars until the gas supply is exhausted
(which has yet to happen for LSB dwarfs) or the gas supply is removed by tidal
stripping (the probably event for cluster dE's to halt star formation).  In this
scenario, the stellar densities between LSB dwarfs and dE's should be similar with
expected slightly higher central densities for dwarf ellipticals owing to a expected
slightly higher SFR.  Figure \ref{mu} displays this comparison using the central
surface brightness of dwarf galaxies based on exponential fits.  The LSB galaxies are
taken from the SPARC sample (where 3.6$\mu$m mags are converted to W1 mags) and
follow the same trend as dE's for absolute luminosity versus central surface
brightness.  The dwarf ellipticals are about 0.5 mags higher in central surface
brightness compared to SPARC LSB's, but have similar central densities compared to
the higher SFR sample from 11HUGS (Lee \etal 2007) and James \etal (2004).  A KS test
rejects the hypothesis that the dE and LSB samples are similar (the p-value was below
1\%).  Thus, the scenario proposed herein is that cluster dwarf ellipticals are
cousins to field LSB dwarfs rather than high SFR BCD's or dIrr's.  Their evolution
was slightly faster in the past than current values for LSB galaxies, an evolution
that was presumingly cut short by entrance into a disruptive cluster environment
where the gas supply was stripped and SF was halted.  This scenario makes it
difficult to assemble normal ellipticals from dwarf ellipticals, as their stellar
populations have very different components, and will be a challenge for galaxy
formation scenarios that depend on dry mergers.

\medskip

\noindent Acknowledgements: 

The author gratefully acknowledges numerous conversations with my SPARC colleagues,
Stacy McGaugh and Federico Lelli.  This project was motivated by discussions with
Nelson Caldwell in the early 1980's.  The software and funding for this project was
supported by NASA's Applied Information Systems Research (AISR) and Astrophysics Data
Analysis Program (ADAP) programs.  Data used for this study was based on observations
made with (1) the NASA Galaxy Evolution Explorer, GALEX is operated for NASA by the
California Institute of Technology under NASA contract NAS5-98034, (2) SDSS where
funding has been provided by the Alfred P.  Sloan Foundation, the Participating
Institutions, the National Science Foundation, the U.S. Department of Energy, the
National Aeronautics and Space Administration, the Japanese Monbukagakusho, the Max
Planck Society, and the Higher Education Funding Council for England, (3) the
Wide-field Infrared Survey Explorer (WISE), which is a joint project of the
University of California, Los Angeles, and the Jet Propulsion Laboratory/California
Institute of Technology, funded by the National Aeronautics and Space Administration
and (4) archival data obtained with the Spitzer Space Telescope, which is operated by
the Jet Propulsion Laboratory, California Institute of Technology under a contract
with NASA.  In addition, this research has made use of the NASA/IPAC Extragalactic
Database (NED) which is operated by the Jet Propulsion Laboratory, California
Institute of Technology, under contract with the National Aeronautics and Space
Administration. 

\pagebreak

\end{document}